\documentclass[
 reprint,
 amsmath,amssymb,
 aps,
]{revtex4-1}
\usepackage{graphicx}
\usepackage{dcolumn}
\usepackage{bm}
\usepackage{braket}
\begin{document}
\title{
Spectroscopy of pionic atoms 
in $\mathbf{{}^{122}{\textbf Sn}({\textit d},{}^3{\textbf He})}$ reaction
and angular dependence of the formation cross sections
}
\author{T.~Nishi$^{1,2}$}
\author{K.~Itahashi$^2$}
\email{Email: itahashi@riken.jp}
\author{G.P.A.~Berg$^{3}$}
\author{H.~Fujioka$^{4}$}
\author{N.~Fukuda$^2$}
\author{N.~Fukunishi$^2$}
\author{H.~Geissel$^{5}$}
\author{R.S.~Hayano$^1$}
\author{S.~Hirenzaki$^{6}$}
\author{K.~Ichikawa$^1$}
\author{N.~Ikeno$^{7}$}
\author{N.~Inabe$^2$}
\author{S.~Itoh$^1$}
\author{M.~Iwasaki$^2$}
\author{D.~Kameda$^2$}
\author{S.~Kawase$^{8}$}
\author{T.~Kubo$^2$}
\author{K.~Kusaka$^2$}
\author{H.~Matsubara$^{8}$}
\author{S.~Michimasa$^{8}$}
\author{K.~Miki$^{1}$}
\author{G.~Mishima$^{1}$}
\author{H.~Miya$^2$}
\author{H.~Nagahiro$^6$}
\author{M.~Nakamura$^2$}
\author{S.~Noji$^1$}
\author{K.~Okochi$^1$}
\author{S.~Ota$^{8}$}
\author{N.~Sakamoto$^2$}
\author{K.~Suzuki$^{9}$}
\author{H.~Takeda$^2$}
\author{Y.K.~Tanaka$^{1}$}
\author{K.~Todoroki$^1$}
\author{K.~Tsukada$^{2}$}
\author{T.~Uesaka$^2$}
\author{Y.N.~Watanabe$^{1}$}
\author{H.~Weick$^{5}$}
\author{H.~Yamakami$^{4}$}
\author{K.~Yoshida$^2$}

\affiliation{\it {}$^1$ Department of Physics, School of Science, The University of Tokyo, 7-3-1 Hongo, Bunkyo-ku, 113-0033 Tokyo, Japan}
\affiliation{\it {}$^2$ Nishina Center for Accelerator-Based Science, RIKEN, 2-1 Hirosawa, Wako, 351-0198 Saitama, Japan}
\affiliation{\it {}$^3$ Department of Physics and the Joint Institute for Nuclear Astrophysics Center for the Evolution of the Elements, 
University of Notre Dame, Notre Dame, Indiana 46556, USA}
\affiliation{\it {}$^4$ Department of Physics, Kyoto University, Kitashirakawa-Oiwakecho, Sakyo-ku, Kyoto, 606-8502 Kyoto, Japan}
\affiliation{\it {}$^5$ GSI Helmholtzzentrum f\"{u}r Schwerionenforschung GmbH, Planckstrasse 1, D-64291 Darmstadt, Germany}
\affiliation{\it {}$^6$ Department of Physics, Nara Women's University, Kita-Uoya Nishimachi, Nara, 630-8506 Nara, Japan}
\affiliation{\it {}$^7$ Department of Life and Environmental Agricultural Sciences,
Faculty of Agriculture, Tottori University, 4-101 Koyamacho-Minami, Tottori, 680-8551 Tottori, Japan}
\affiliation{\it {}$^{8}$ Center for Nuclear Study, The University of Tokyo, 2-1 Hirosawa, Wako, 351-0198 Saitama, Japan}
\affiliation{\it {}$^{9}$ Stefan Meyer Institute for Subatomic Physics, Austrian Academy of Sciences, Boltzmanngasse 3, A-1090 Vienna, Austria}
\collaboration{piAF Collaboration}

\date{\today}

\begin{abstract}
We observed the atomic $1s$ and $2p$ states of $\pi^-$ bound to ${}^{121}{\rm Sn}$ nuclei
as distinct peak structures in the missing mass spectra of the
${}^{122}{\rm Sn}(d,{}^3{\rm He})$ nuclear reaction.
A very intense deuteron beam and a spectrometer with a large angular
acceptance let us achieve potential of discovery, which 
includes capability of determining the angle-dependent cross sections with high statistics.
The $2p$ state in a Sn nucleus was observed for the first time.
The binding energies and widths of the pionic states are determined and found
to be consistent with previous experimental results of other Sn isotopes.
The spectrum is measured at finite reaction angles for the first time.
The formation cross sections at the reaction angles between 0 and $2^\circ$ are determined.
The observed reaction-angle dependence of each state is reproduced by theoretical calculations.
However, the quantitative comparison with our high-precision data reveals a
significant discrepancy between the measured and calculated formation cross sections
of the pionic $1s$ state.
\end{abstract}


\maketitle

The spectroscopy of pionic atoms has 
contributed to the fundamental knowledge of the non-trivial structure of the
vacuum in terms of chiral symmetry and
quantum chromodynamics (QCD) in the low energy region~\cite{Kienle04}.
The spatial overlaps between the pionic orbitals and the core nuclei
peak near the half-density radii of the nuclei. 
The pions are excellent probes for the study of medium effects in the nuclear matter.
An order parameter of the chiral symmetry~\cite{Gell_Mann68,Tomozawa66,Weinberg66},
a quark condensate expectation value $\Braket{\bar{q}q}$, was deduced at the 
nuclear density 
in investigations on in-medium modification of an isovector $\pi$-nucleon 
strong interaction through the wave
function renormalization~\cite{Kolomeitsev03,Suzuki04,Hayano10,Yamazaki12,Friedman07,Friedman14}.

The low energy pion-nucleus interaction
is described by a phenomenological optical potential.
Parameter sets of the potential were obtained 
by fitting many known pionic-atom
and pion-nucleus-scattering data
including isotope shifts of pionic atoms with different neutron numbers~\cite{Batty97,Yamazaki12,Friedman07,Friedman14,Konijn90}.
Among these parameters is an $s$-wave isovector parameter $b_1$, which
is closely related to the order parameter of the chiral 
symmetry breaking in the nuclear medium $\Braket{\bar{q}q}_\rho$~\cite{Kolomeitsev03}. 
Low-lying pionic orbitals are located in a 
close vicinity of the core nuclei for $Z \gtrsim 50$~\cite{Toki89}.
This localized distribution is due to 
integration of the attractive Coulomb potential and the repulsive and absorptive
strong interaction potential. Determining the levels and widths of the bound states
provides quantum-mechanical information that leads to constraints on the
strong interaction.
Previous experiments discovered methods of directly populating
the low-lying orbitals, analyzed them spectroscopically~\cite{Yamazaki96,Gilg00,Itahashi00}, and
measured pionic states in Pb and Sn nuclei. The
ratio of $\Braket{\bar{q}q}_\rho$ to the in-vacuum value $\Braket{\bar{q}q}_0$ was evaluated to be
$\Braket{\bar{q}q}_{\rho}/\Braket{\bar{q}q}_0 \sim 67\% $~\cite{Suzuki04}
based on the in-medium modification of the isovector interaction, which is in good agreement with
chiral perturbation theories~\cite{Meissner02}.

Measurements of the pionic atom formation cross sections at finite
reaction angles provide unprecedented information.
This contributes
to a better understanding of the pionic-atom formation mechanisms
and leads to higher accuracy in deduction of the fundamental quantities.
Experimental data at larger reaction angles open
new prospects for the identifications of the
quantum numbers corresponding to the structures
observed in the spectra. Previously one had merely to rely on the
comparison between the measured and theoretical spectral 
shapes~\cite{Itahashi00}. 
Up to the present experiment, the formation cross sections were measured only in
very limited solid angles around $0^\circ$. Meanwhile, theories 
predict reaction-angle dependence
mainly originating from the
different momentum transfers in the $(d,{}^3{\rm He})$ reactions~\cite{Ikeno15}.
Only a small angular dependence is expected
from the elementary $\pi^-$ production cross sections of the
$n(d,{}^3{\rm He}) \pi^-$ reaction~\cite{Frank54}.

\begin{figure}[hbtp]
 \includegraphics[width=8.3cm]{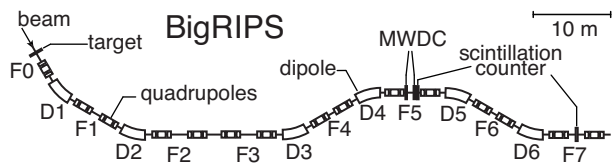}
 \caption{A schematic view of the experimental facility,
   BigRIPS~\cite{Kubo03}, and the detector setup.
   A deuteron beam impinged on a target located at the nominal target position F0.
   The emitted ${}^3{\rm He}$ particles were momentum-analyzed at the dispersive focal plane at F5
and the tracks were measured by the MWDCs. The scintillation counters at F5 and F7 were
used for particle identification.}
\label{Fig:1}
\end{figure}

We conducted spectroscopic measurements of
the missing mass of the ${}^{122}{\rm Sn}(d,{}^{3}{\rm He})$ nuclear reactions
at the RI Beam Factory RIBF, RIKEN~\cite{Yano07} in October 2010.
The missing mass of the ${}^{122}{\rm Sn}(d,{}^3{\rm He})$ reaction near the pion
emission threshold was measured for the first time.
The experiment demonstrated the excellent performance of
RIBF applied
in an experimental program of high precision spectroscopy with the
primary beam.
We employed a deuteron beam
with a typical intensity of $2 \times 10^{11}$/s
accelerated
by the cyclotron complex of AVF-RRC-SRC and focused onto the target location of 
the BigRIPS in-flight magnetic separator~\cite{Kubo03}.
The beam was extracted in micro bunches of frequency of 13.7 MHz
and had a high duty factor.
The beam energy has been determined to be $T_d = 498.9 \pm 0.2$ MeV 
by NMR measurements in BigRIPS.
The horizontal beam emittance and intrinsic momentum spread have been
estimated to be $\sim 0.54 \times 3.0\ \pi$ mm$\cdot$mrad ($\sigma$)
and $0.04 ^{+0.01}_{-0.02}\%$, respectively.

Figure \ref{Fig:1} illustrates the schematic layout of 
the employed detectors and BigRIPS used as a spectrometer.
The acceptance aperture of $\sim 2^\circ$ was utilized to cover a 
reaction-angle range centered near $0^\circ$.
The deuteron beam
impinged on a 1 mm wide strip target
of ${}^{122}{\rm Sn}$ at F0 which was isotopically enriched to
95.8\% and had a thickness of $12.5\pm0.5$ mg/cm$^2$.
The achieved mean luminosities of about $10^{31}$ cm$^{-2}$ s$^{-1}$
in the present experiment were much higher than those of
previous experiments at GSI~\cite{Suzuki04,Yamazaki96,Gilg00,Itahashi00} despite of the thinner target.
We expected direct formation of pionic atoms coupled with neutron hole states
in the nuclear reactions near the recoil-free kinematical condition.
Relevant neutron hole states (excitation energies $E_n(n'j'l')$) in the ${}^{121}{\rm Sn}$ nucleus
are
$2d_{3/2}$ (0.0 MeV), 
$3s_{1/2}$ (0.06034 MeV), 
$2d_{5/2}$ (1.1212 MeV), and
$2d_{5/2}$ (1.4035 MeV)~\cite{Ohya10}.

The emitted ${}^3{\rm He}$ particles were momentum analyzed at F5
with a momentum dispersion of
$60.64\pm 0.15$ mm/\%. We installed two sets
of multi-wire drift chambers (MWDC) 
near the momentum dispersive focal plane at F5
and measured the tracks of the charged particles with
a tracking resolution of $\sim$ 40 $\mu$m ($\sigma$).
The ${}^3{\rm He}$ ions were identified
by the time-of-flight of $\sim 174$ ns
between F5 and F7 measured by the scintillation
counters.
The counters had a distance of about $23$ m.
We achieved almost background-free ${}^3{\rm He}$
spectra ($< 0.1 $ \% contamination).
A typical
trigger rate
of the data acquisition was 200/s due to 
the relatively narrow coincidence gate between F5 and F7 detectors.
The deadtime of the data-acquisition system has been estimated to achieve an
efficiency of $\sim 93\%$.

The measured ${}^3{\rm He}$ tracks have been used
to determine the ${}^3{\rm He}$ kinetic energies,
the positions, and the angles at the target after detailed
analyses and corrections of the optical transfer coefficients up to the
fifth order aberrations~\cite{Nishi13}.
Hereafter measured ${}^3{\rm He}$ emission 
angles are treated as the reaction angle $\theta$. The $\theta$ resolution has been
estimated to be $\sim 0.3^\circ$ ($\sigma$).
The mass of the reaction product, the pionic atom, has been
determined by the measured ${}^3{\rm He}$ energy 
and $\theta$. It has been related to the excitation energy $E_{\rm ex}$ 
relative to the ground-state mass of ${}^{121}{\rm Sn}$ ($M({}^{121}{\rm Sn}$)).
The corresponding relation at $0^\circ$ is given by the equation
$E_{\rm ex} = [M_{\rm mm} - M({}^{121}{\rm Sn})]c^2
    = m_{\pi^-}c^2 - B_{nl} + E_{\rm n}(n'l'j')$ 
where $M_{\rm mm}$ is the mass of the reaction product,
$B_{nl}$ is the binding energy of the pion 
in a state characterized by the principal ($n$) and angular ($l$) quantum numbers.
$m_{\pi^-}=139.571$~MeV$/c^2$ is the $\pi^-$ rest mass.

The ${}^3{\rm He}$ kinetic energy has been calibrated by using the two-body reaction of
$p(d,{}^{3}{\rm He})\pi^0$. The target consists of a 
$100 \pm 1~\mu$m thick 2~mm wide polyethylene strip. We performed
a calibration measurement every two hours during the production
runs. The calibration spectra have low-energy tails
due to the large reaction angles. Therefore, we have examined the spectral response
by Monte Carlo simulations including effects of matter, spectrometer
acceptance and detector performances.
A fit has been conducted to reproduce the entire spectral shape in the plane
of the measured positions and angles and related
them to the ${}^{3}{\rm He}$ energies. The systematic errors
of the absolute $E_{\rm ex}$ values
have been estimated to be $^{+0.036}_{-0.033}$~MeV ($\sigma$).

The $p(d,{}^{3}{\rm He})\pi^0$ reaction has also been used to calibrate 
the effective beam intensity on the target.
The reaction cross section has been estimated to be 7.6~$\mu$b/sr
in the $\theta$ range of 0 -- 0.5$^\circ$
based on an extrapolation of the data 
at a slightly higher energy to ours~\cite{Chapman64}.
For this we have used the measured beam energy dependence in Ref.~\cite{Betigeri01}.
After applying an acceptance correction of the spectrometer evaluated by
a Monte Carlo simulation, we have estimated a systematic error for the
absolute cross section scale of 30\%.

The experimental resolution has been estimated 
by the quadratic sum of contributions from the incident beam emittance and intrinsic
momentum spread, the target thickness, and the optical aberrations of the spectrometer.
We have found a quadratic dependence of the resolution on $E_{\rm ex}$ due to
combined effects of multiple scattering at a vacuum window at F5
and higher-order optical-aberrations. This dependence has been estimated to be
$R(E_{\rm ex}) = \sqrt{R_{\rm min}^2 + (0.122\times(E_{\rm ex} - 139.799~{\rm MeV}))^2}$(FWHM)
with the resolution minimum $R_{\rm min} = 0.42$ MeV, which 
agrees with the measured spectral responses in the calibration reaction of $p(d,{}^{3}{\rm He})\pi^0$.

\begin{figure}[hbtp]
  \includegraphics[width=8.7cm]{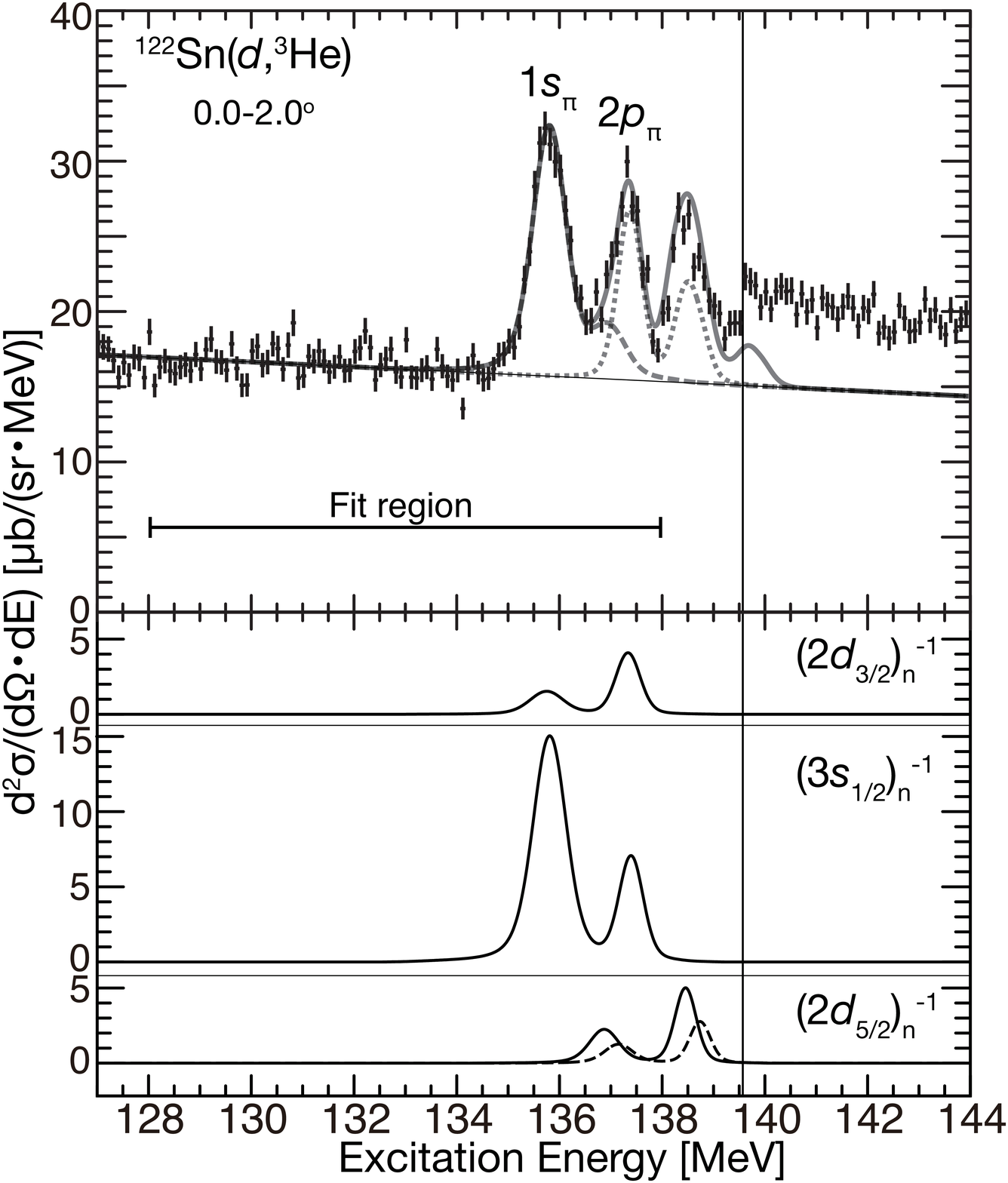}
  \caption{(Top panel) Measured excitation spectrum of the ${}^{122}{\rm Sn}(d,{}^3{\rm He})$ reaction
    at the angular range of $0<\theta<2^\circ$.
Three distinct peaks are observed in the region $E_{\rm ex} = $
[134,139] MeV. The left and middle
peaks are mainly originating from
formation of pionic $1s$ and $2p$ states, respectively.
The right peak is partly contributed from the other pionic states
$(2s, 3p,$ and $3s)$.
The spectrum is fitted in the region indicated.
The fitting curve and contributions from the $1s$ and the $2p$
states are presented
by solid, dashed and dotted lines, respectively.
(Bottom panel) Decomposition of the pionic $1s$ and $2p$ strengths into contributions from each neutron hole state
of ${}^{121}{\rm Sn}$ as indicated. Note that fragmentation is taken into consideration for
$(2d_{5/2})^{-1}_n$.
}
\label{Fig:2}
\end{figure}

Figure~\ref{Fig:2} (top panel) displays the measured excitation spectrum
for nearly the full acceptance of the spectrometer.
The abscissa is the excitation energy
and the ordinate is the double differential cross section of the
${}^{122}{\rm Sn}(d,{}^3{\rm He})$ reaction.
The $\pi^-$ emission threshold is indicated by the vertical solid line at 139.571~MeV.
On the left side of the spectrum in the range of $E_{\rm ex} <$ 134 MeV, a
linear background is observed for nuclear excitation without 
pion production. Above the emission threshold a continuum is observed due to quasi-free pion production. 

Three prominent peaks are observed below the pion emission threshold 
in the region of 134 MeV $\lesssim E_{\rm ex} \lesssim$ 139 MeV.
The leftmost peak is due to 
the formation of a pionic $1s$ state mainly coupled with a neutron 
hole state of $(3s_{1/2})_n^{-1}$. 
The middle peak contains contributions from the configurations $(1s)_\pi(2d_{5/2})_n^{-1}$, 
$(2p)_\pi(3s_{1/2})_n^{-1}$ and $(2p)_\pi(2d_{3/2})_n^{-1}$.
The peak on the right side
originates mainly from the $(2s)_\pi(3s_{1/2})_n^{-1}$ and $(2p)_\pi(2d_{5/2})_n^{-1}$
configurations.

The spectrum has been fitted in an excitation energy region 
[128.0,138.0]~MeV
with calculated spectra based on
theoretical pionic atom formation cross sections
in Ref.~\cite{Ikeno11} folded by
the experimental resolution expressed by Gaussian functions.
Pionic $1s$, $2p$, $2s$, $3p$, and $3s$ states have been taken into considerations and
other higher states as well as the quasi-free contributions have been neglected. 
In the fit 8 parameters have been used: the differential cross sections ($d\sigma/d\Omega$) of
pionic $(nl)$ states $I_{1s}, I_{2p}$, the binding energies $B_{1s}, B_{2p}$, 
the $1s$ width $\Gamma_{1s}$, and a slope and an offset for the linear background.
The $2p$ width has been fixed to a calculated value of 0.109~MeV~\cite{Ikeno15,Ikeno16p} since it is
much smaller than the experimental resolution.
Since contributions from the other states $2s$, $3p$ and $3s$
are small, their binding energies and widths have been fixed to theoretical
values and their relative cross section ratios $I_{2s}/I_{1s}$, $I_{3s}/I_{1s}$ and $I_{3p}/I_{2p}$
to theoretical ratios~\cite{Ikeno15,Ikeno16p}.
The resolution minimum $R_{\rm min}$ has also been used as a free parameter.

The fitted curve is presented with contributions from the pionic $1s$ and $2p$ 
states. The overall fit has a $\chi^2$/n.d.f of 135.8/92. 
Figure~\ref{Fig:2} (bottom panel) shows the decomposition of the $1s$ and $2p$
formation cross sections into different neutron hole
states of ${}^{121}{\rm Sn}$ as indicated.
The peak on the left is coupled with pionic $1s$ and the one on the right with $2p$.

We have evaluated the systematic errors attributed to the deduced binding energies and width
resulting from
i)~absolute $E_{\rm ex}$ scale error arising from the energy calibration, the
uncertainty of the primary beam energy, the uncertainty of the target thickness
and the ion-optical properties of the spectrometer,
ii) the $E_{\rm ex}$ dependence of the resolution within evaluated errors,
iii)~the fitting region and iv)~20\% errors in the spectroscopic factors of relevant neutron holes.
The systematic errors of the binding energies are mainly arising 
from the energy calibration and the dispersion of the spectrometer.

The binding energies and width are deduced to be
\begin{eqnarray*}
B_{1s} &=& 3.828\pm 0.013(stat.) ^{+0.036}_{-0.033}(syst.) {\rm ~MeV} \\
\Gamma_{1s} &=& 0.252\pm 0.054(stat.)^{+0.053}_{-0.070}(syst.) {\rm ~MeV}\\
B_{2p} &=& 2.238\pm 0.015(stat.)^{+0.046}_{-0.043}(syst.) {\rm ~MeV}.
\end{eqnarray*}
with the statistical and systematic errors. The achieved resolution minimum of $R_{\rm min} = 0.394^{+0.064}_{-0.044}$ MeV
is consistent with the estimation of 0.42~MeV described above.
$B_{1s}$ and $B_{2p}$ of a pionic Sn isotope are determined simultaneously for the first time.
The deduced $B_{1s}$, $\Gamma_{1s}$ and $B_{2p}$ fairly well agree with the theoretical values
$B^{\rm theo}_{1s}=3.787$ -- $3.850$ MeV, $\Gamma^{\rm theo}_{1s}=0.306$ -- $0.324$ MeV and $B^{\rm theo}_{2p}=2.257$ -- $2.276$ MeV~\cite{Ikeno16p,Ikeno15}.

\begin{figure}[hbtp]
\includegraphics[width=8.7cm]{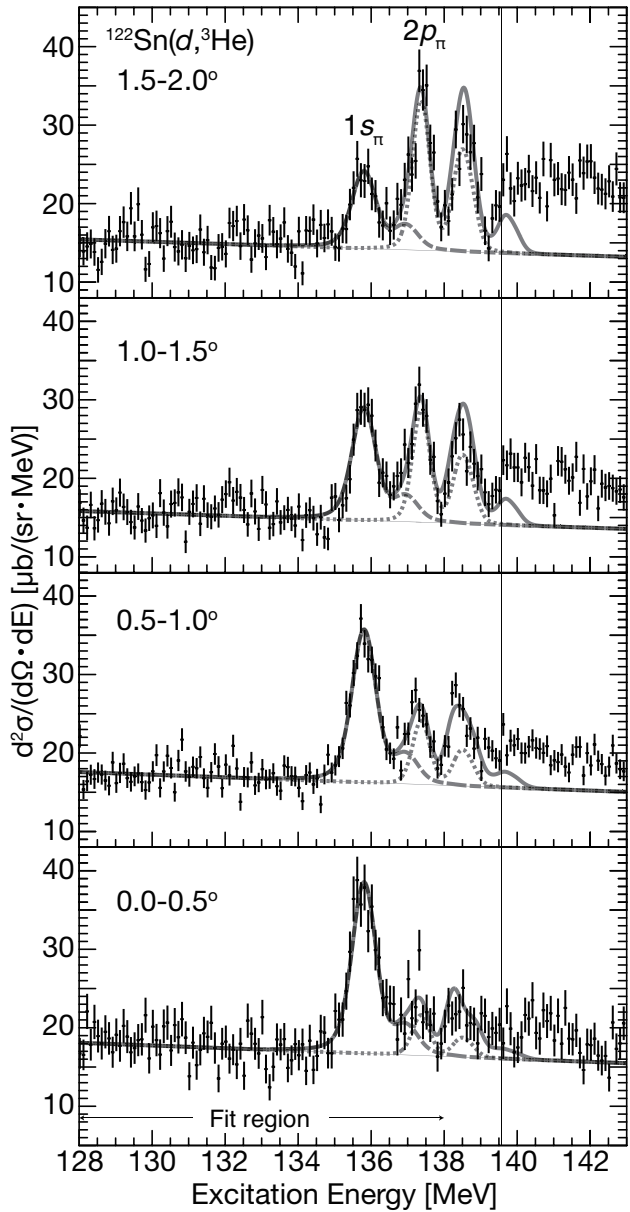}
\caption{The measured excitation spectrum in Fig.~\ref{Fig:2} 
is decomposed into spectra with different
$\theta$ ranges as indicated.
The overall fitting curves and the $1s$ and $2p$ components are 
shown by the solid, dashed and dotted lines, respectively.}
\label{Fig:3}
\end{figure}

The measured spectrum has been decomposed into different $\theta$ ranges
expecting the $\theta$ dependence of the formation
cross sections~\cite{Ikeno11,Ikeno11A,Ikeno15}. 
Figure~\ref{Fig:3} shows the excitation spectra
for $\theta$ in the ranges of 1.5 -- 2.0$^\circ$, 1.0 -- 1.5$^\circ$,
0.5 -- 1.0$^\circ$ and 0 -- 0.5$^\circ$.
The peak structures are clearly observed
for each $\theta$ range. The peak positions and widths are nearly constant for
the different $\theta$ ranges, which demonstrates the
high quality of the measurement.

The decomposed spectra have been fitted to deduce $\theta$-dependent
cross sections for the pionic $1s$ and $2p$ states, $I_{1s}(\theta)$ and $I_{2p}(\theta)$,
with other parameters fixed to those determined in the fitting procedure described above.
The linear background has been scaled by a parameter for each angular range.
The resulting fitting curves are shown together with curves for $1s$ and $2p$ contributions.

\begin{figure}[hbtp]
\includegraphics[width=7.8cm]{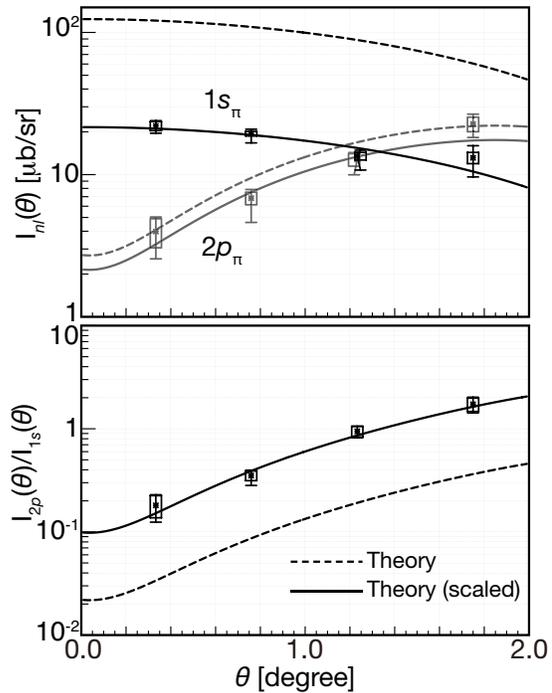}
\caption{(Top panel) Determined pionic-$nl$-state formation cross sections $I_{nl}(\theta)$
  for different $\theta$ ranges. 
  Statistical errors are shown by the boxes
  and systematic errors in addition by the bars. 
  The deduced cross sections are compared with the theoretical calculations~\cite{Ikeno15,Ikeno16p}.
  (Bottom panel) $I_{2p}(\theta)/I_{1s}(\theta)$. Systematic errors are canceled by taking the ratios.
}
\label{Fig:4}
\end{figure}

Figure~\ref{Fig:4} (top panel) depicts $I_{1s}(\theta)$ (black) and $I_{2p}(\theta)$ (grey).
The abscissa is $\theta$
represented by the weighted averages of the solid angles.
The boxes show the statistical errors and the bars the systematic errors in addition.
They have been studied in the same way as those conducted for the binding energies and widths.  
An overall systematic error of 30\% is attributed to the absolute scale of the cross section.
The dashed curves show theoretically calculated cross sections of the pionic $1s$ and $2p$ states
using phenomenological neutron wave functions of Koura-type given in Ref.~\cite{Koura00}.
The solid curves show the same but scaled with
fitted factors of 0.17 and 0.79 for the $1s$ and $2p$ states, respectively.
We observe that the theoretical $\theta$ dependences reproduce well the experimental data,
which confirms the assignments of the angular momenta of the states.
In principle, this suggests that the theoretical models of finite
momentum transfer~\cite{Ikeno15} are valid. 
However, the absolute value of
measured $I_{1s}(\theta)$ is much smaller than the theoretical calculation.
Figure~\ref{Fig:4} (bottom panel) shows the ratios of the $I_{2p}(\theta)$ and $I_{1s}(\theta)$.
The large discrepancies between the experimental data and the theory
suggest missing factors of the formation cross sections
which have to be independently applied to the pionic states.

In conclusion, we have performed 
spectroscopy of
pionic ${}^{121}{\rm Sn}$ atoms
and observed the $1s$ and $2p$ states as prominent peak structures.
For Sn, the $2p$ state is 
observed as a peak structure for the first time.
We have determined the binding energies of the $1s$ and $2p$ states
and the width of the $1s$ state.
The reaction-angle dependences of the pionic atom formation cross sections are measured
and found to agree with the theoretical dependences, which
supports the assignments of the quantum-numbers of the measured peak structures.
We also find remarkable agreement with theoretically calculated $2p$ formation cross sections
over a wide range of reaction angles. However, the measured absolute 
cross sections of the $1s$ state are smaller by a factor of $\sim$ 5.
Note that the entire data were accumulated within 15 h,
which is showing the potential of the facility RIBF
for spectroscopy experiments.
Continuous development is in progress aiming at 
a better spectral resolution of $\leq 150 $ keV.
The major accomplishments in the present experiment will be succeeded by
experiments with improved resolution, statistics and systematics errors to
deduce $\pi$-nucleus isovector scattering length $b_1$ with better
accuracy~\cite{Itahashi08}.
A new series of experiments to study pionic atoms over
a wide range of nuclei is in preparation and will
lead to a better understanding of the fundamental structure
of the QCD vacuum based on measurements.

The authors thank Prof. Emeritus Dr.~Toshimitsu Yamazaki for
fruitful discussions and
late Prof. Emeritus Dr.~Paul~Kienle for his guidance in this study.
The authors are grateful to the
staffs of GSI for providing target materials
and staffs of RIBF for stable operation of the facility.
This experiment was performed at RI Beam Factory operated by RIKEN Nishina Center and CNS,
University of Tokyo.
This work is partly supported
by MEXT Grants-in-Aid for Scientific Research on Innovative Areas
(Grants No. JP22105517, No. JP24105712 and No. JP15H00844), JSPS
Grants-in-Aid for Scientific Research (B) (Grant No. JP16340083), 
(A) (Grant No. JP16H02197) and (C) (Grants No. JP20540273 and No. JP24540274),
Grant-in-Aid for JSPS Research Fellow (Grant No. 12J08538),
the Bundesministerium f\"ur Bildung und Forschung,
and the National Science Foundation through Grant No.
Phys-0758100, and the Joint Institute for Nuclear Astrophysics
through Grants No. Phys-0822648 and No. PHY-1430152 (JINA Center for the
Evolution of the Elements).

\end{document}